\documentclass[fleqn,usenatbib]{mnras}

\usepackage{newtxtext,newtxmath}
\usepackage{soul}
\usepackage{ulem}

\usepackage[T1]{fontenc}

\DeclareRobustCommand{\VAN}[3]{#2}
\let\VANthebibliography\thebibliography
\def\thebibliography{\DeclareRobustCommand{\VAN}[3]{##3}\VANthebibliography}

\usepackage{graphicx}	
\usepackage{amsmath}	

\numberwithin{equation}{section}

\defcitealias{Pokhrel+2021}{PGK21}
\defcitealias{Palau+15}{P15}
\defcitealias{Palau+21}{P21}
\defcitealias{Lada+10}{LLA10}
\defcitealias{Peltonen+24}{PRW24}
\defcitealias{Grossschedl+19}{G19}

\title[intra-cloud $N_{\rm yso}$-M$_{\rm gas}$ Relation]{Comparing the M$_{\rm gas}$-N$_{\rm yso}$ Relation inside a Giant Molecular Cloud.}

\author[C. G. Rom\'an-Z\'u\~niga et al.]{
Carlos G. Rom\'an-Z\'u\~niga,$^{1}$\thanks{croman@astro.unam.mx}
A. Palau,$^{2}$
J. Ballesteros-Paredes,$^{2}$
M. Zamora-Avil\'es,$^{3}$  
\newauthor{J. Peltonen$^{4}$}
and K. Guti\'errez-D\'avila$^{2}$
\\
$^{1}$Universidad Nacional Aut\'onoma de M\'exico, Instituto de Astronom\'ia, AP 106, Ensenada 22800, BC, M\'exico\\
$^{2}$Universidad Nacional Autónoma de M\'exico, Instituto de Radioastronom\'ia y Astrof\'isica, Antigua Carretera a Patzcuaro 8701,\\  Ex-Hda. San Jos\'e de la Huerta,
58089 Morelia, Michoac\'an, M\'exico\\
$^{3}$Instituto Nacional de Astrof\'isica, \'Optica y Electr\'onica, Luis E. Erro 1, Tonantzintla 72840, Puebla, M\'exico\\
$^{4}$Department of Physics, University of Alberta, Edmonton, AB T6G 2E1, Canada\\
}

\date{Accepted XXX. Received YYY; in original form ZZZ}

\pubyear{\the\year{}}

\begin{document}
\label{firstpage}
\pagerange{\pageref{firstpage}--\pageref{lastpage}}
\maketitle

\begin{abstract}
In this paper we present a simple analysis around scaling relations derived from the Schmidt conjecture for star-forming molecular clouds, at the intra-cloud scale. Using a hierarchical tree (dendrograms) above a constant threshold ($A_V=7$ mag), we separate individual gas structures in a column density map of the nearby Giant Molecular Cloud Orion A, constructed from Herschel far-infrared maps. These structures define regions of dense molecular gas that can actively form stars. We also estimate their current embedded population using a list of known young stars. From the combined analysis of the column density map and the young star catalog, we  construct a series of plots that show the intra-cloud level behavior of three well-known scaling relations: $N_{\rm yso}$ {\it vs.} $M_{\rm gas}$, $\Sigma_{\rm SFR}$ {\it vs.} $\Sigma_{\rm gas}$ and $R_{\rm eq}$ {\it vs.} $M_{\rm gas}$. Our dataset, along with other sets from literature, show the validity of a linear relation for $N_{\rm yso}$ {\it vs.} $M_{\rm gas}$, from intra-cloud to inter-cloud scales, over three orders of magnitude. We also especulate on the possibility that the relation could be valid over an even larger scale range. Additionally, our data are consistent with the $R_{\rm eq}$ {\it vs.} $M_{\rm gas}$ discussed in previous studies. However, our data is not quite in agreement with previously proposed fits for the $\Sigma _{\rm SFR}$ {\it vs.} $\Sigma _{\rm gas}$ relation, and we discuss the implications of using the free-fall timescale as the main parameter defining the star-forming efficiency in dense gas regions. 
 \end{abstract}




\section{Introduction}\label{s:intro}

A formalism for large scale star formation is still absent in modern astrophysics. The phenomenon involves a diversity of very complex processes acting simultaneously, thus requiring one to break the problem into a number of general concepts. The most obvious one is the understanding that Giant Molecular Clouds (GMC) collapse by the action of gravity, and that such collapse leads to star formation, as the densest regions in the cloud are both gathered and segregated\footnote{for a specific discussion on density segregation see \citet{Alfaro_Roman-Zuniga18}}. This idea was first introduced by \citep[][the so called Schmidt's Law]{Schmidt59}, and forms the basis for the validity of the widely invoked Kennicut-Schmidt relation \citep[KSR; ][]{Kennicutt98} from galactic to extragalactic scales \citep{Kennicutt_Evans12}. In the case of Milky Way's molecular clouds, this relation connects two main observables: the surface density of gas, that can be measured from 2-dimensional maps, and the surface density of the young stellar objects (YSO) that we can observe forming within the clouds by directly counting, or by inferring, the number of YSOs present in a star-forming cloud (\citet{Evans+09}; \citet{Gutermuth+11}; \citet[][hereafter L10]{Lada+10}; \citet{Lada13}; \citet{Zari+19}; \citet[][hereafter PGK21]{Pokhrel+2021}, \citet{Spilker+22}). 

Several recent studies have shed light on how the KSR may arise at the disks of galaxies  \citep[see a full discussion in][]{paperI}, but also how the limitations of observational data blur important details: for instance, it is difficult to make complete censuses of all stars forming within a GMC, and it is also complicated to define the relevant structures of clouds; at extragalactic scales, in the best cases we resolve individual GMCs, but not yet the regions in those GMCs that are forming stars. Moreover, the parental gas is immediately affected by massive stars as they emerge \citep[see e.g.][]{Leisawitz+89, Ballesteros-Paredes_Hartmann07, Roman-Zuniga+15}, separating the parental gas from the YSOs and from the gas ionized by them. This effect, while irrelevant for unresolved extragalactic star-forming regions, it is crucial for \textit{inter-cloud} (comparison among clouds) and \textit{intra-cloud} (comparison of structures within clouds) studies where both gas and stellar populations are resolved \citep{Kruijssen+14,Kruijssen+19,Pessa+21}. In the work of \citetalias{Lada+10}, it was shown that when considering all the gas mass detected by a morphologically coherent tracer \citep[e.g. dust extinction, dust emission, CO integrated intensity;][]{Ridge+06} it is not possible to establish a relation to the total number of YSO. Instead, a clear correlation appears when only considering the mass of \textit{dense gas}, defined above a proper threshold. Recent work by \citet{jiao25} and \citet{rawat25} confirm this trend in a collection of both low-mass and high-mass star forming clouds. Similarly, the work by \citet{Palau+24} confirms that this trend holds when considering only the lowest mass population of the clouds.

In the study of \citetalias{Pokhrel+2021}, they studied the intra-cloud KSR in a dozen nearby ($\mathrm{d<1.5\ kpc}$) GMCs, finding a power-law correlation with index 2 between the surface density of star formation, $\Sigma_{\rm SFR}$, and the surface density of gas, $\Sigma_{\rm gas}$. Then they showed that the correlation is linear if $\Sigma_{\rm gas}$ is normalized by the free-fall time scale $(\tau_{\rm ff})$. They also estimated the star-forming efficiency per free-fall timescale, $\epsilon _{\rm ff}$ \citep{KruMcK+05}, and proposed that is approximately constant within the clouds and that varies around an average of $0.03$ within less of 1 dex variation among clouds.  In~ \citet{paperI} \ it was proposed that $\epsilon _{\rm ff}$ can be better understood as the ratio between the star formation rate and the gas infall rate during the star formation process. These authors also proposed that the simplest version of the KSR, relating a mass of star forming gas and the number of stars formed in it (the $\mathrm{N_{\rm yso}-M_{\rm gas}}$ relation) can be defined as:
\begin{equation}
    \mathrm{ N_{\rm yso}=\left( \frac{\epsilon_{\rm ff} \langle \tau_{\rm yso}\rangle}{\langle m_{\rm yso}\rangle \tau_{\rm ff}}\right )M_{\rm gas}}
    \label{e15p1}
\end{equation}
where $\epsilon_{\rm ff}$ is the efficiency per free-fall time ($\tau_{\rm ff}$) and $\langle m_{\rm yso}\rangle,\ \langle\tau_{\rm yso}\rangle$ account for the typical mass and age of the young stars.

In \citet[][]{paperII} using simulated observations of star-forming clouds, it was shown that values of $\epsilon _{\rm ff}$ comparable to those of \citetalias{Pokhrel+2021} are indeed common to large portions of the clouds, even though local star formation efficiencies, directly measured from the simulation could be as high as 30 percent. They proposed how small, constant values of  $\epsilon _{\rm ff}$ are partly a consequence of using YSO lifetime scales that sometimes differ from $\tau_{\rm ff}$, which is derived from supposing spherical radial profiles for the star-forming structures. 

It is important to consider that several previous studies determine the surface density of star formation $\Sigma _{\rm SFR}$ on counting YSOs projected above column density levels in large regions across entire clouds. For the inter-cloud case, a common threshold level is used to define the transition to dense gas that succumb to star formation (e.g $A_V=7$ mag in \citetalias{Lada+10}, \citet{Lada13}), and the whole area of a cloud above the threshold level is integrated along with all possible YSO counts as a single data-point. For the intra-cloud case, the idea is to  consider that, as the column density level is increased, the area of the contributing regions is reduced, but each level can be fragmented into more sub-structures as long as the Jeans length admits \citep[e.g. ][]{Roman-Zuniga+10, Palau+18}. A few studies used all individual fragments (e.g. density peaks) within a given level \citep{Pokhrel+20, Roman-Zuniga+2019}. In this paper we make use of a column density map and a YSO catalog for the Orion A GMC. We use dendrograms to consider all relevant intra-cloud structures and to carefully count the projected sources within those structures. We will show how some scale relations are clearly kept for an evolved massive region like Orion A. We also discuss the reasons and consequences of the differences we observe in the context of the previous two studies in this series.

The paper is organized as follows: In section \ref{s:datamethods} we describe the different datasets and our analysis; in section \ref{s:results} we describe the main results obtained from our analysis; we discuss those results in section \ref{s:discsum}, along with a summary of the contents of the paper in section \ref{s:sum}.

\section{Datasets and Methodology} \label{s:datamethods}

In this section we describe the datasets we used, as well as the methods applied to them for the purposes of our analysis. We combine a high quality column density map of the Orion A GMC along with the catalog by \citet[][hereafter G19]{Grossschedl+19}, which is considered one of the most complete YSO catalogs of the complex collected to date. We used their classification by evolutionary class, allowing us to provide a mean age and a simple estimate of stellar mass. We also made careful counts of the projected number of YSOs contained within the projected areas of relevant structures (dense regions) in the GMC.

\begin{figure*}
    \centering
    \includegraphics[width=2.1\columnwidth]{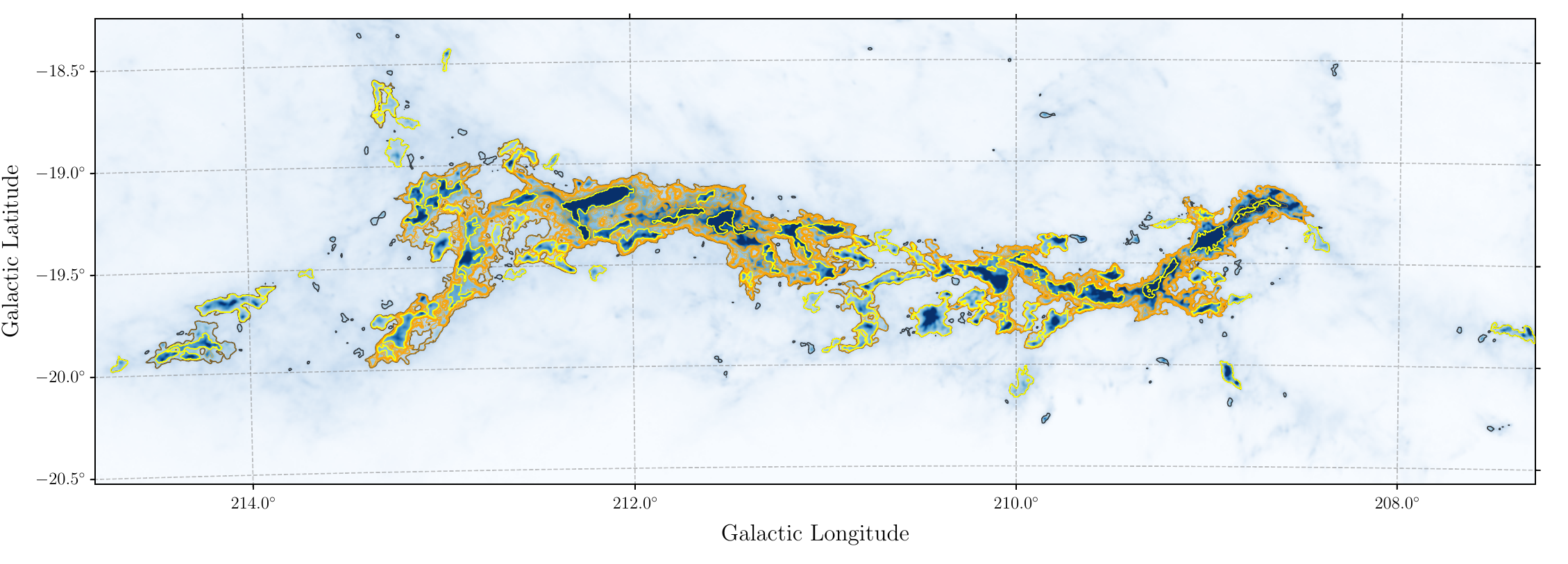}
    \caption{HP2 extinction map of the Orion A Cloud. The grayscale shows extinction values from 0.5 to 20 mag in logarithmic scale. The red colored contour delineates the $A_V=7$ mag level on the map. The orange colored contours delineate the boundaries of branch type objects. The yellow colored contours delineate the boundaries of leaf type objects.}
    \label{fig:OrionGLB}
\end{figure*}

\subsection{Column Density maps } \label{s:datamethods:ss:avmaps}

We made use of the Orion A Herschel–Planck–2MASS (HP2) column density map, described in the work of \cite{Lombardi+14}. The Orion A complex offers a wide range of star forming activity, from relatively quiescent to highly active star forming regions. 

In a nutshell, the HP2 methodology is based on the combination of thermal dust emission images from the PACS (60 and 100 $\mathrm{\mu m}$) and SPIRE (250, 350 and 500 $\mathrm{\mu m}$) bolometers, obtained as part of the Gould Belt Survey \citep{Andre+10} of the Herschel Observatory. They used the spectral energy distributions for the Herschel instruments passbands at each position in the maps to fit the dust temperature at each point. They used for this purpose the dust emissivity, estimated from the \cite{Planck+11} maps of the same regions, that is calculated with a spatial resolution of 35 arcsec, similar to the resolution of the SPIRE 500 $\mathrm{\mu m}$ to which the rest of the images are convolved to. Then, a photometric absolute flux calibration was applied by using the lower resolution Planck dust temperature maps \footnote{the nominal resolution of the Planck dust temperature maps is 5 arcmin}. This way they were able to obtain precise opacities at 850 $\mathrm{\mu m}$ which, in turn, were finally converted to dust extinction $\mathrm{A_K}$ using a conversion factor that is obtained directly from a linear fit to 2MASS near-infrared excess \citep[NICEST;][]{Lombardi+09} maps they constructed for the same areas. The nominal (beam) resolution of the maps, $b_{sz}$ is 30 arcsec in the HP2 map.

\subsection{Orion A VISION YSO Catalog \label{s:datamethods:ss:vision}}

We made use of the Orion A YSO catalog constructed by \citetalias{Grossschedl+19}. That catalog was derived from a list of almost 6$\times 10^3$ objects detected in the near-infrared (J,H and K$_s$ bands) VISTA Survey of Orion A \citep[VISION;][]{Meingast+16}, and contains a total 2980 individual objects catalogued as candidate young star members of the Orion A Complex. The VISION catalog is, to date, one of the most complete inventories of young stars in the Orion A Complex. We separated out from the list those objects catalogued as non-resolved or ambiguous, as well as those catalogued as having high probabilities of being background galaxies and quasars. The VISION catalog also contains columns with the quasar and galaxy probabilities from the Data Source Catalog Combined Module \citep{Jamal24}. In Appendix \ref{App.1} we briefly discuss why the VISION classification works better than Gaia's for the purpose of this work. Our final list contains a total of 2801 objects catalogued as Class I (131 sources), Class II (167 sources; 298 together with Class I) and Class III (1650 sources) YSOs, as well as objects catalogued as having transitional and debris disks (650 objects; we considered these as Class III for our analysis). This means that we considered, for the purposes of this study, a majority of the objects formed in the last 5 Myr within the cloud, including those formed in large clusters like the ONC at the northern Integral Shaped Filament, and smaller clusters like those in the L1641 region at the South end of the complex (for a recent discussion on cluster content of the Orion A Complex, see \citet{SSJ24}). Notice that \citet{Pokhrel+20} used a list of 294 Class I and 2100 Class II YSOs in Orion A, which suggest that their Class I classification could be a combination of the Class I and Class II classification by \citetalias{Grossschedl+19}, while Class II in \citet{Pokhrel+20} may contain a large portion of sources defined as Class III, transition disks and disk in remission in the work of \citetalias{Grossschedl+19}.
 
\subsection{Relevant Gas Structures from Dendrograms \label{s:datamethods:ss:dendro}}

In previous studies we employed a methodology to find individual density peaks across column density maps. Such methodology is useful when making statistics of global parameters, determining the significance of sub-structures \citep{Roman-Zuniga+09,Roman-Zuniga+10,Roman-Zuniga+2019}, but it fails to provide a system to organize the structures in hierarchies. In addition, for the nature of our analysis, we need to consider star forming activity considering regions of dense gas within the clouds, but not necessarily the activity in each possible substructure. Hence, in this work we decompose the 2-dimensional images using the Python package \texttt{astrodendro} \citep{Rosolowsky+08}, which allows us to decompose the maps in nested, hierarchical structures called dendrograms.

For all our dendrogram runs we set $\mathrm{\Delta_{min}}=0.5$ mag, corresponding in our case to the minimum height difference for structures to be retained within a level. We also required that structures in the dendrogram trees to encompass a minimum of 100 pixels to be considered as significant. We constructed dendrogram trees for all our maps using an increasing minimum level, $\mathrm{min(A_V)}$, from 1.0 to 10.0 magnitudes. However, for this paper we only considered the trees with $\mathrm{min(A_V)}=7.0$ mag ($\mathrm{min(A_K)}\approx 0.8$ mag) that, as discussed in \cite{Lada13} contains over 80 percent of protostars in Orion A, and thus defines a simple and clear criterion for dense gas structures with active formation. 

The objects derived from dendrograms \texttt{astrodendro} are catallogued as ``trunks'' (t), ``branches'' (b) and ``leaves'' (l), depending on whether a structure is split in more pieces or not. For simplicity, in some cases we call generically to all these structures as ``clumps'', and use the abreviation ``cl''. The\  \texttt{astrodendro} algorithm also allows to define groups of leaves belonging to the same trunk or "ancestor". 

We found a total of 156 objects in the Orion A H2P map, from which 62 are classified as "branches" and 94 as "leaves". These structures are divided into 4 ancestor groups, which we named G1 to G4. Groups G2 and G3 are the most populated, with 54 and 36 structures, respectively, and correspond roughly to the L1641 South region and the Integral Shaped Filament in the Orion A Complex. Groups G1 and G4 contain only 3 and 2 structures each, and are small regions located in the cloudlets of the Whale Tail, East of $\mathrm{l}=214^\circ$. In 98 of the objects we found more than 3 YSO sources projected within their areas (see section \ref{s:datamethods:ss:clumpprops}). Objects with less than 3 or no YSO counts, were discarded from the analysis.

In Figure \ref{fig:OrionGLB} we show the Orion A extinction map along with the locations the extensions of the dendrogram objects constructed from \texttt{astrodendro} contours.

\subsection{Star-forming and Physical Properties of Clumps \label{s:datamethods:ss:clumpprops}}

\begin{figure}
    \centering
    \includegraphics[width=\columnwidth]{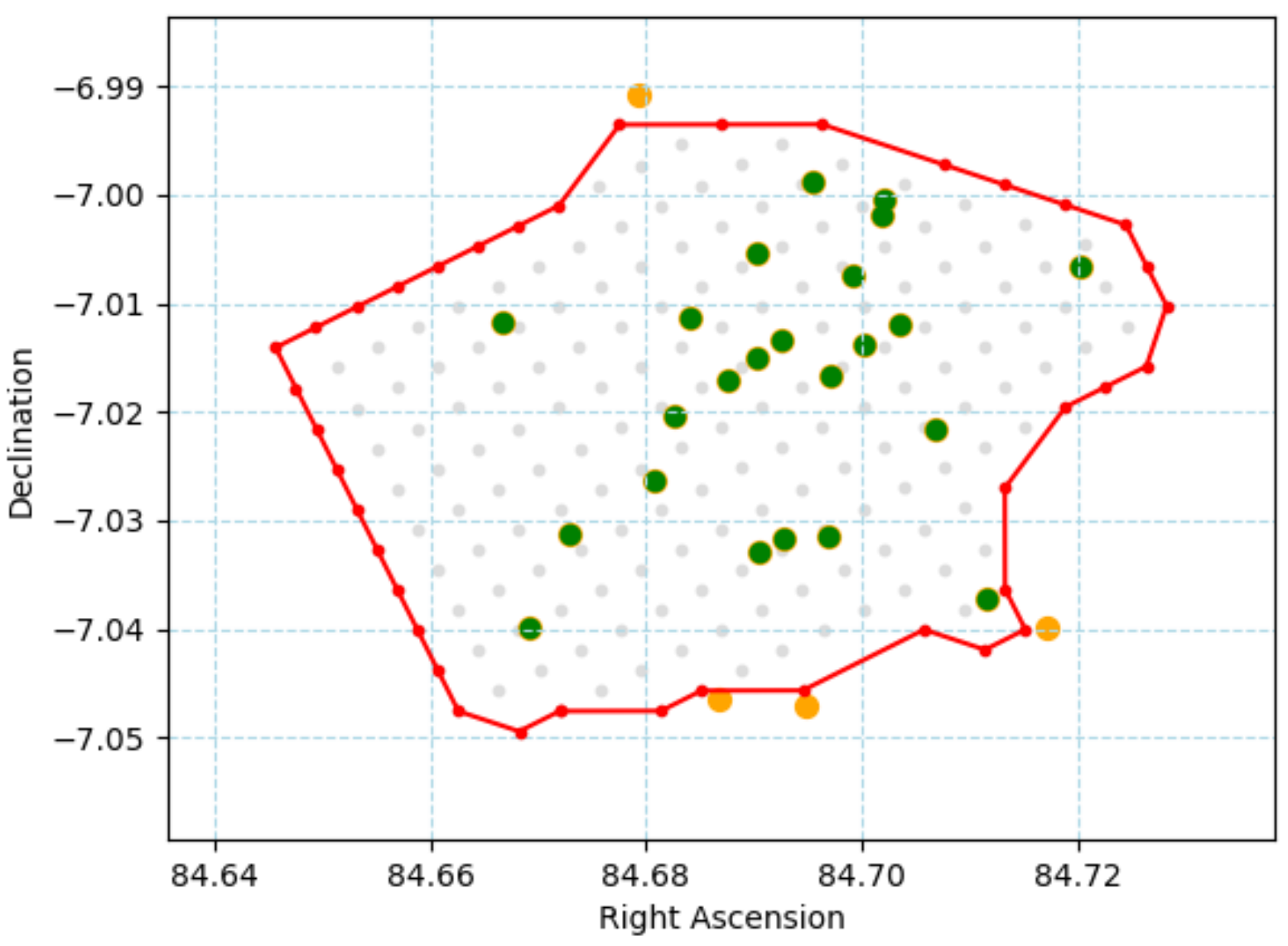}
    \caption{The image illustrates how a group of YSO positions is matched inside the boundaries of a pixel structure in one of our maps. Gray dots trace the pixel positions of a clump in the extinction map. The red dots and line indicate the positions of pixels used to define the polygon boundary of the structure. The green dots are objects falling inside the boundary. Orange dots are rejected as they do not fall completely inside the boundary.}
    \label{fig:shapely}
\end{figure}

One of our goals was to measure the star formation rate in the clumps. The 2-dimensional morphology of the dense gas regions is highly irregular, and the stars are also projected in very irregular patterns over these areas. We used the contours from \texttt{astrodendro} at $\mathrm{A_V}=7.0$ mag to determine the polygon borders and the pixel areas enclosed by them using the python package \texttt{Shapely}\footnote{\href{https://pypi.org/project/shapely}{https://pypi.org/project/shapely} \citep{shapely2007}. \texttt{Shapely} is in turn based on the open source GEOS library \citep{geos}.}. We defined the irregular borders of clumps as concave hull shapes defined by the pixel groups assigned to them. By combining with the astronomical coordinate routines of the \texttt{Astropy} package, we determined precisely which YSO positions were located inside the object borders above the threshold level using world coordinates. It is important to notice that a number of YSOs are located outside the threshold border, and we made sure those cases were not accounted. We show an example in Fig. \ref{fig:shapely}. Our methodology, while not infallible, allows us to estimate, as closely as possible, the number of \textit{embedded} YSOs within the gas clumps defined by the objects borders defined with \texttt{astrodendro}. 

\subsubsection{Age and Mass for YSOs}\label{s:datamethods:ss:clumpprops:sss:ysoagemass}

For each one of the YSO catalogs, we made a rough estimate of the age and mass of the sources. Using the listed evolutionary class (available for all the catalogs we compiled from the literature described above), we assigned ages of 0.5, 2.0 and 3.0 Myr to YSOs of class I, II and III, respectively. Regarding the Class I ages, the age of 0.5 Myr is a good agreement with previous studies, particularly with \citet{Palau+15} and \citet{Palau+21}, whose data is included in our analysis. About our choice for Class II and Class III ages, the age separation takes into consideration typical ages for T Tauri stars and considers that while the timescale for disk dissipation can be rapid (~0.5 Myr), there appears to be a consensus on a well defined, age difference between classes, which we compromise to be at least 1 Myr. We include a more detailed justification in Appendix \ref{App.3}. Still, considering that the scatter around mean ages can be very significant, we considered a 50\% uncertainty in the ages of stars for subsequent calculations.

\begin{figure}
    \centering
    \includegraphics[width=1.0\columnwidth]{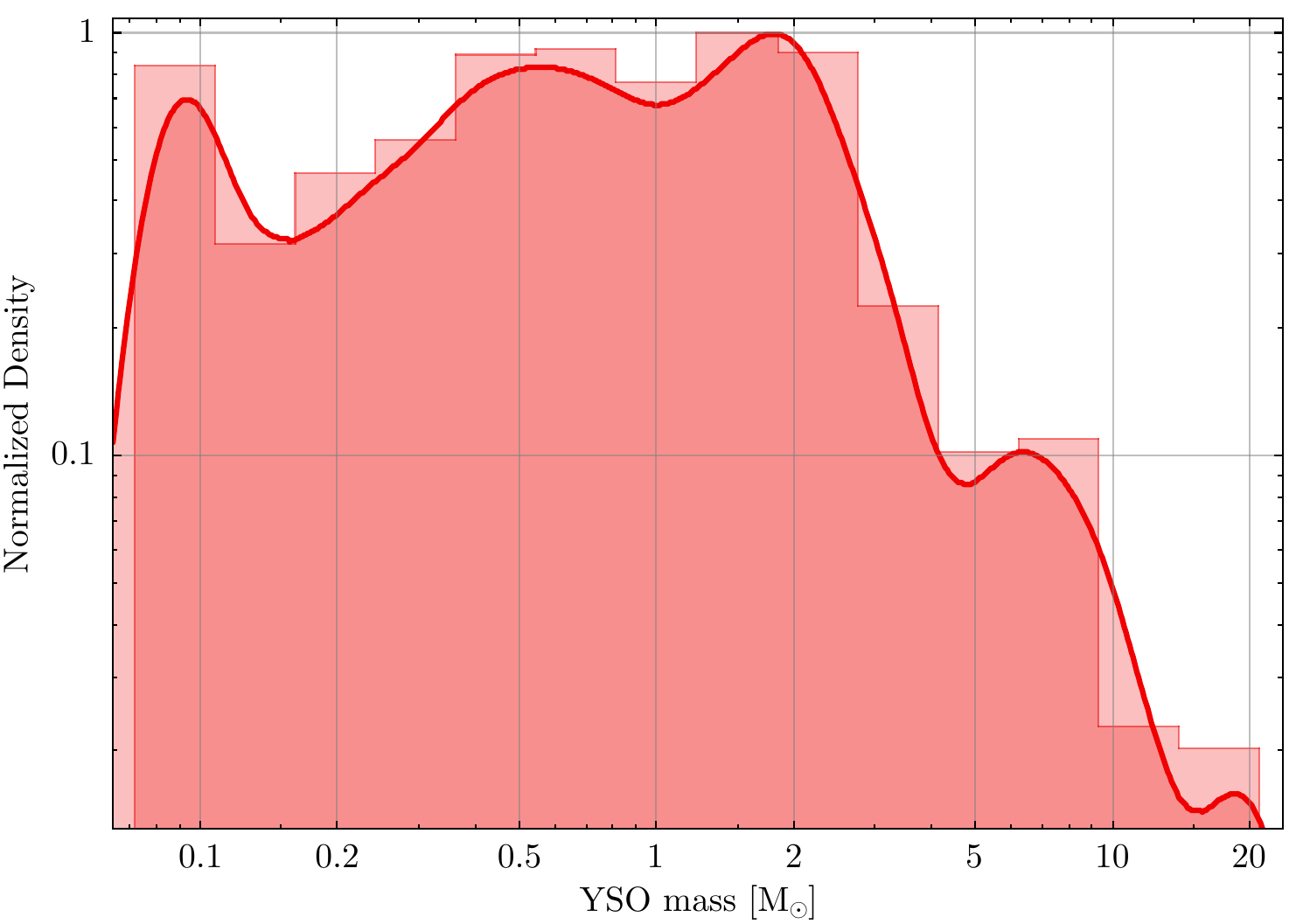}
    \caption{The histogram shows the distribution of YSO mass values estimated by the method described in section \ref{s:datamethods:ss:clumpprops:sss:ysoagemass}. The solid red line is a Kernel Density Estimate of the same distribution made with a 3$\sigma$ truncated Gaussian.}
    \label{fig:ysomasshist}
\end{figure}

In order to assign masses to individual YSOs, we used corresponding isochrone models for 0.5, 2.0 and 3.0 Myr in PARSEC library \citep{PARSEC} to interpolate mass values. We used the listed $A_V$ values in the catalog of \citetalias{Grossschedl+19} to have de-reddened $K_s$ (2.12 $\mu$m) band magnitude and then applied a distance modulus correction using Gaia DR3 parallaxes. In Figure \ref{fig:ysomasshist} we show a histogram of the YSO mass values obtained through this procedure. 

One of the goals of the paper is to consider how the efficiency in a star forming structure is a balance between the gas available for collapse and the removal of gas by the stars formed. Since it is possible that a fraction of evolved YSOs (older Class II and Class III sources) are partially dispersed outside the structure, this could seriously affect the counts of truly embedded YSOs, specially for small structures. While an accurate estimation of this effect is not possible with these data, we took it into consideration in two ways: First, considering that a young star moving away from its birthplace at a speed of 1 km/s would travel around 1 pc in one Myr, we removed from the analysis those objects with equivalent radii below 0.3 pc. Second, we made our estimates of surface density of star formation using, in one case all YSOs and in another case using only the Class I+II source counts.

Moreover \citet{Lombardi+13,Lombardi+14} discussed that it is safe to consider the dispersion of young stars away from their birth sites as a minor effect in $N_{\rm yso}$ counts. Also, in the study of \citet{Ybarra+13}
it was shown that while star formation progresses across a cloud towards the regions of higher density, the Class II and Class III form layers that stay reasonably close to the $A_v=8$~mag boundaries, which is consistent with our suppositions. It is also important to point that gas dispersal timescale relevant for scale relations, considering relatively massive clusters like the ONC, is about 5 Myr \citep{Zhou+25}, and possibly longer in regions absent of massive stars like L1641.

\subsubsection{Properties of Gas Clumps}\label{s:datamethods:ss:clumpprops:sss:gasprops}


For each of the clumps we estimated their area ($A_{\rm cl}$ in units of sq. pc) equivalent radii, gas mass $M_{\rm gas}$ and surface gas densities, $\Sigma_{\rm gas}$, directly from the object catalogs and the extinction maps:

\begin{equation}
R_{\rm eq}=(A_{\rm cl}/\pi)^{1/2}\ [\mathrm{pc}],
\label{e2.1}
\end{equation}

 \begin{equation}
M_{\rm gas}=1.28\times 10^{-10} d_{\rm cl}^2 b_{sz}^2 \frac{1}{4}\sum _{i}A_{V,i}\ [M_\odot],
\label{e2.2}
 \end{equation}

\begin{equation}
\Sigma_{\rm gas}=M_{\rm gas}/A_{\rm cl} \ [M_\odot\cdot pc^{-2}],
\label{e2.3}
\end{equation}
where the sum in equation \ref{e2.2} is made on the extinction values of all pixels enclosed within the clump border, and $d_{\rm cl}$ is the distance to the clump. In the case of Orion A, we used data from Table 2 of \citetalias{Grossschedl+19} which lists mean distances to YSOs from the VISION catalog as a function of Galactic longitude ($l$). By fitting a 4th order polynomial to those data, we were able to assign a distance to the central position of each clump in our object catalog  by using their Galactic longitude coordinate.

Also, following the prescription of \citetalias{Pokhrel+2021}, we made estimates for the gas volume density in each clump, $\rho_{\rm cl}$, which allows us in turn to estimate the "free-fall" or cloud collapse timescale \citep{Krumholz+12}:

\begin{equation}
\rho_{\rm cl} = \frac{3(\pi)^{1/2} M_{\rm gas}}{4A_{\rm cl}^{3/2}} \ \ [M_\odot/\mathrm{pc}^3]
\label{e2.4}
\end{equation}

\begin{equation}
\tau_{\rm ff,cl} = \left( \frac{3\pi}{32\cdot 4.492\times 10^{-9} \rho_{\rm cl}} \right)^{1/2} \ \ [{\rm Myr}]    
\label{e2.5}
\end{equation}

With the addition of the YSO catalogs we estimated the number and total mass of YSOs enclosed in each structure, allowing us to estimate, for each clump, a star-forming rate, $\mathrm{SFR}$ and a surface density of star formation, $\Sigma_{\rm SFR}$, which we estimate using the mass ($m$) and age ($\tau$) of YSOs defined above for each evolutionary class (I, II or III):

\begin{equation}
\mathrm{SFR} = \sum _{i=1}^{N_{I}} \frac{m_{i,I}}{0.5} + \sum _{i=1}^{N_{II}} \frac{m_{i,II}}{2.0} + \sum _{i=1}^{N_{III}} \frac{m_{i,III}}{3.0} \ \ [\mathrm{M_\odot \cdot Myr^{-1}}].
\label{e2.6}
\end{equation}

Once the SFR is calculated, we used it to estimate the surface density of star formation as:

\begin{equation}
 \Sigma _{\rm SFR} = \frac{SFR}{A_{\rm cl}} \ \  [\mathrm{M_\odot\cdot Myr^{-1}\cdot pc^2}] 
 \label{e2.7}
\end{equation}

This in turn allows the estimation of the star-forming efficiency per free-fall time, $\epsilon_{\rm ff}$ \citep{Krumholz+12, Pokhrel+2021}:

\begin{equation}
\epsilon_{\rm ff}= \frac{\mathrm{SFR}}{M_{\rm gas}/ \tau_{\rm ff,cl}}.
\label{e2.8}
\end{equation}

Uncertainties for all parameters were calculated as follows: using the opacity error map in the HP2 files, we converted to extinction and estimated the total of the uncertainty for each object. We also estimated the uncertainty in the area of each object by considering an average value of 3 percent error in the determination of the distances to each structure. We estimated Poisson errors for the YSO counts and we used a constant 10 percent error for the YSO mass values, considering that the precision of our interpolation using the evolutionary models is mainly dependent on photometric errors of at few tenths of magnitudes at most. Using these considerations, we then estimated uncertainties in all calculations using standard error propagation formulas.

It is worth commenting that while our method to estimate the SFR values improves in detail over the use of single (constant) values for mass and age of YSOs (see also section \ref{s:discsum}), it is still a rough estimate; for instance, we are using evolutionary class definitions from the \citetalias{Grossschedl+19} catalog, that may differ from other authors. Also, the ages and masses that can be obtained from the PARSEC isochrones may differ slightly from those that could be obtained from another isochrone model set. 

\section{Results} \label{s:results}

The main results for this study involve several scale relationships at the intra-cloud level, derived from the analysis of the Orion A data, along with data from \citetalias{Palau+15} and \citetalias{Palau+21} (hereafter P15, and P21, respectively).

\subsection{Number of YSOs versus Clump Gas Mass \label{s:results:ss:nysomass}}

\begin{figure}
    \centering
    \includegraphics[width=\columnwidth]{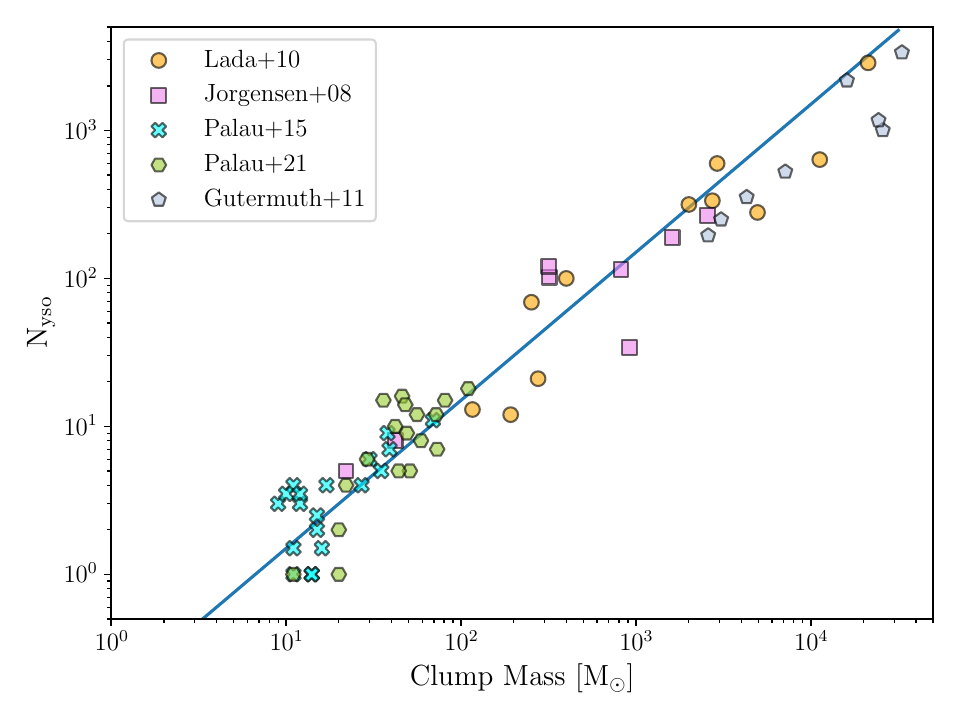}
    \caption{ $N_{\rm yso}$ {\it vs.} $M_{\rm gas}$ relationship as derived for the inter-cloud studies of \citet{Lada+10} and \citet{Gutermuth+11}, the intra-cloud study of \citet{Jorgensen+08} and the individual clump determinations in the studies of \citetalias{Palau+15} and \citetalias{Palau+21}. The blue line shows a general fit to the data of the form $N_{\rm yso}=0.15 \cdot M_{\rm gas}^{1.0}$.}
    \label{fig:McNyso_lit}
\end{figure}

In Figure \ref{fig:McNyso_lit} we show the relationship between the number of YSOs and cloud mass at three different scales, from different literature data. The first dataset is from \citetalias{Lada+10}, where the total mass of gas and the total number of protostars contained in nine Giant Molecular Clouds above a column density threshold are considered. They used extinction maps constructed with the NICEST method \citep{Lombardi+09} and protostar counts, which were obtained from Spitzer catalogs from the C2D program\footnote{\href{http://irsa.ipac.caltech.edu/data/SPITZER/C2D/}{http://irsa.ipac.caltech.edu/data/SPITZER/C2D/}}. 
In the study of \citet{Lombardi+14}\footnote{see Fig. 19 in their corrigendum, \citet{Lombardi+14C}}, they show that their Orion A HP2 map holds a mass over $\sim 2.1\times 10^4\rm M_{\odot}$ above $A_K=0.8\ A_V\sim 7$, 1.55 times larger than the $\sim 1.3\times 10^4\rm M_{\odot}$ reported above the same extinction level by \citet{Lada+10} for the NICEST map. This is probably due to the fact that the Herschel data probes higher extinction levels than 2MASS. We suppose this correction to be valid for all points in the \citet{Lada+10} table, and we multiplied all cloud dense gas mass values by a factor of 1.55.

The second dataset belongs to the work of \cite{Jorgensen+08}, which contains gas mass measurements from millimetric continuum peaks in two clouds observed with JCMT/SCUBA. They also used the C2D catalogs to estimate the number of protostars in the clumps defined by the continuum peaks. The third and fourth datasets belong to the studies of \citet{Palau+15} and \citet{Palau+21} (hereafter P15, and P21, respectively), which are based on estimates of dense core masses and numbers of millimetre peaks in each core, associated with the presence of protostars. The fifth dataset is from the inter-cloud study of \citet{Gutermuth+11} who also used near-IR extinction maps and YSO catalogs from a number of studies involving Spitzer and Chandra space telescope observations; for this case we added the counts for Class I and Class II sources together. 

There is a very tight agreement between the different datasets. It is worth noticing that, at face value, the datasets from \citetalias{Palau+15} and \citetalias{Palau+21} extend the $M_{\rm gas}$ {\it vs.} $N_{\rm yso}$ correlation down to clump masses near 1 M$_\odot$. The general relation is well adjusted with a function of the form $N_{\rm yso}=A\cdot M_{\rm gas}^{\alpha}$ with $\alpha=1.0$ and $A$=0.15, in good agreement with the slope of $0.96\pm 0.11$ found for the inter-cloud case in \citetalias{Lada+10}.

This confirms a linear $N_{\rm yso}$ {\it vs.} $M_{\rm gas}$ relation for both the inter-cloud and the intra-cloud regime (as in the works of \citetalias{Lada+10} and \cite{Jorgensen+08}), as well as for individual clump measurements of \citetalias{Palau+15} and \citetalias{Palau+21}. This suggests that the $N_{\rm yso}$ {\it vs.} $M_{\rm gas}$ relation is valid over at least 3 orders of magnitude in cloud mass.

\begin{figure}
    \centering
    \includegraphics[width=\columnwidth]{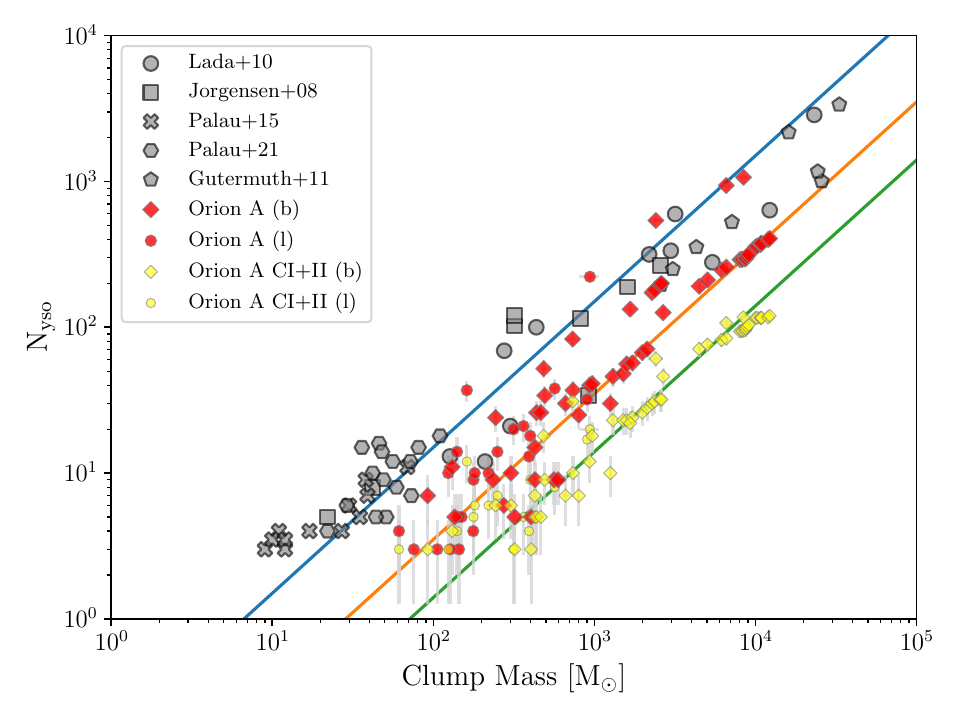}
    \caption{$N_{\rm yso}$ {\it vs.} $M_{\rm gas}$ relation for the intra-cloud dataset presented in this study. The red symbols in two shapes (circles and diamonds) separate branches and leaves in the A$_V>7.0$ dendrogram object list obtained from the Orion A map. The yellow symbols represent counts with only Class I and II sources. We show, for comparison, the same literature datasets as in Fig. \ref{fig:McNyso_lit} respecting the same symbols but using a single color. The solid lines are linear best fits for the data groups $(y=Ax;\ A={0.15, 0.035, 0.014})$.}
    \label{fig:McNyso_main}
\end{figure}

In Fig.~\ref{fig:McNyso_main} we again plot the number of YSOs vs the clump mass, this time adding the data of Orion A, presenting both the counts for Class I and II (yellow points), and Class I to III (red points) sources. We notice that each case can be fitted with a linear relation. The differences in the intercept are due to the fact that, when substracting the Class III sources, the countings in each structure are lower. However, it is interesting to notice that each case, with some scatter, exhibits a linear relation. Our fits for each case are $y=0.035x$ and $y=0.014x$.

The apparent lack of a agreement in the $y-$intercept for the Orion A data with those of Palau, Jorgensen and Lada samples is a consequence of the differences in the inclusion of younger or older YSOs and considering different density thresholds in the mass of dense gas, which in turn correspond to a different $\rm \tau_{yso}$ and $\tau_\mathrm{ff}$ in eq.~\eqref{e15p1}, as we discuss in $\S$\ref{s:discsum}.

 We now explored the possibility that the $\mathrm{N_{\rm yso}~{\it vs.}~M_{\rm gas}}$ relation could be valid at a larger range of spatial scales. In a recent study, \citet[][hereafter PRW24]{Peltonen+24} showed a $N_{\rm yso}$ {\it vs.} $M_{\rm gas}$ for the inter-cloud case but using a collection of clouds observed in the galaxy M33 (their Figure 7). Their estimates for $N_{\rm yso}$ is inferred from an extrapolation of the  Initial Mass Function  based on the detection of massive stars with JWST/MIRI, and the cloud mass values come from $^{12}$CO(2-1) data obtained as part of the ALMA/ACA survey. 
 
\begin{figure}
    \centering
    \includegraphics[width=\columnwidth]{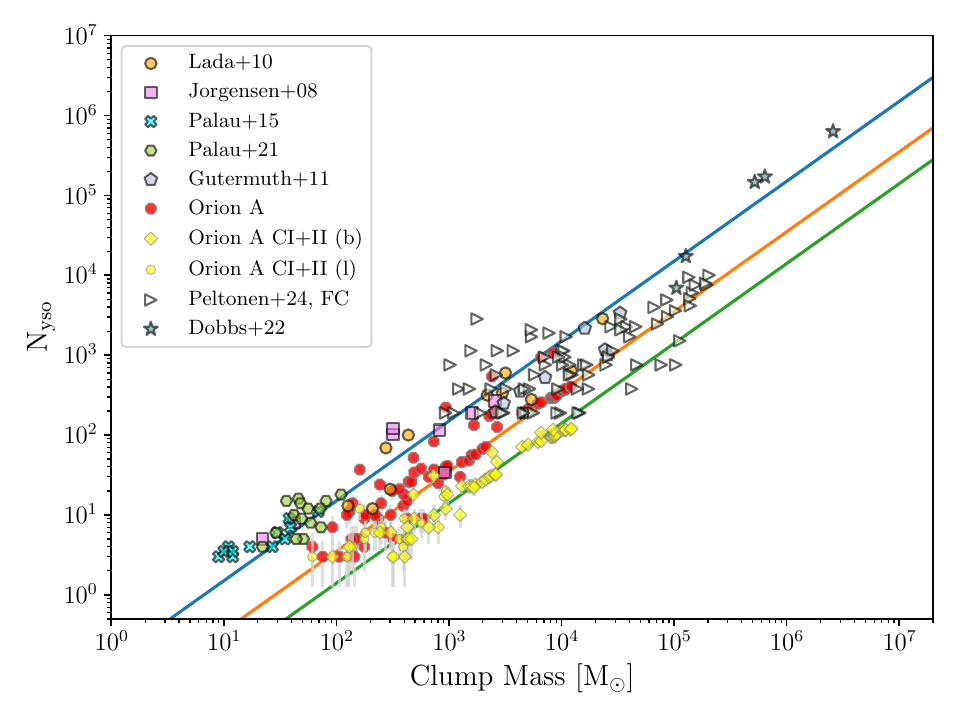}
    \caption{This plot combines $N_{\rm yso}$ {\it vs.} $M_{\rm gas}$ relations from  figures \ref{fig:McNyso_lit} and \ref{fig:McNyso_main}, along with data from the inter-cloud extragalactic study of \citetalias{Peltonen+24} and the numerical study of \citet{Dobbs+22}. See text for details}
    \label{fig:McMyso_3cD}
\end{figure}

In Fig. \ref{fig:McMyso_3cD} we show the $N_{\rm yso}$ {\it vs.} $M_{\rm gas}$ as computed by Peltonen, but applying corrrections to a) include low mass stars, and b) to consider only dense gas. For (a) we followed  \citetalias{Peltonen+24} and corrected their observed counts $N_{\rm yso}$ that are sensitive above 6 M$_\odot$, and used the Kroupa initial mass function \citep{Kroupa01}, to correct the YSO counts to $\sim 0.08\mathrm{M_\odot}$, similar to the \citetalias{Grossschedl+19} catalog.  For (b) we notice that in Table 2 of  LLA10, the fraction of dense to total gas mass for their cloud list ranges from 2 to 23 percent, with a median value near 10 percent. Similarly, \citet{Zak+25} estimate dense gas mass fractions within molecular clouds (intra-cloud) using 3 mm line emission maps, and find values between 5 and 25 percent, consistent with LLA10. Thus, we consider safe to propose a 10 percent scaling for the dense to total gas mass in the PRW24 clouds. By applying both scalings to the data of PRW24, we obtain acceptable fit for their data. At the lower end of cloud mass the data points present larger scatter, but the agreement between $10^4<M_{\rm gas}<10^5\ \mathrm{M}_\odot$ is remarkably good.

 
In Fig. \ref{fig:McMyso_3cD} we also include an estimation of $N_{\rm yso}$ {\it vs.} $M_{\rm gas}$ from the study of \citet{Dobbs+22}, who simulated the formation of stellar clusters on a spiral arm scale in a Milky Way type galaxy. They did not estimate individual stellar mass for their simulated populations and assumed that the mean stellar mass is 1 M$_\odot$. Still we see that this data points provide a set fairly comparable to our data in the $10^5<M_{\rm gas}<10^7\ \mathrm{M}_\odot$ range. The agreement with the intercept would still be acceptable if the $\mathrm{N_{\rm yso}}$ values changed by 50 percent (which is possible considering the use of a mass distribution instead of a single value). Again, the important point is the linearity observed in the points. We also want to mention that \citet{Zak+25} find in their study a strong linear correlation between the dense gas mass and the intensity of 24 $\mu$m emission (a strong YSO signature), which they claim to be in line with the \citetalias{Lada+10} $N_{\rm yso}$ {\it vs.} $M_{\rm gas}$ relation, valid in the $10^4<M_{\rm gas}<10^6\ \mathrm{M}_\odot$ range, similar to the one of \citeauthor{Dobbs+22}.
 
 At face value, the correlation in each dataset is approximately linear despite the differences among each dataset and the scales involved. This suggests the possibility that the $N_{\rm yso}$ {\it vs.} $M_{\rm gas}$ is valid at all natural scales of star formation in galaxy disks. As we discuss in section 4, this may be related to a similarity in the efficiency of star formation at the distinct scales that involve the conversion of gas into stars.

\subsection{Surface density of Star Formation versus Surface Density of Gas \label{s:results:ss:sfrsgas}}

The study of \citetalias{Pokhrel+2021} proposed that the intra-cloud version of the KS relation is valid in twelve nearby molecular clouds. They used a methodology where they define a valid column density range that encloses the whole projected distribution of the dense gas in the maps they studied, and they derived $\Sigma_{\rm SFR}$ and $\Sigma_{\rm gas}$ by estimating the gas mass and the number of young stars contained at each column density level. Their main results are that a) all clouds obey an intra-cloud relation of the form $\log{\Sigma_{\rm SFR}}=2.0\log{\Sigma_{\rm gas}}-4.11$ within 2-$\sigma$; b) all clouds follow an intra-cloud relation of the form $\log{\Sigma_{\rm SFR}}=0.94\log{(\Sigma_{\rm gas}/\tau_{\rm ff})}-1.53$ within 2-$\sigma$, and c) all clouds are consistent with relatively constant values of $\log{\epsilon_{\rm ff}}$ that differ little from a median of -1.59. The main differences in our case are that we reject those objects with equivalent radii smaller than 0.3 pc (see  $\S$\ref{s:datamethods:ss:clumpprops:sss:ysoagemass}), and those that contain less than 3 stars. Also, each object is considered as a single data point. In contrast, \citetalias{Pokhrel+2021} considered different column density thresholds, and at each threshold, they count all the mass and area as if it were one single cloud, even though at a given column density threshold, there are isolated regions with no stars but that are also considered in the total mass budget of the cloud at that threshold.

In consequence, at similar column densities, \citetalias{Pokhrel+2021} always count larger areas, masses and number of stars than us, while we fragment our cloud into substructures that we consider individually. These methodological differences have additional implications in the interpretation of the KS relation, which we defer for the next paper in this series (K. Gutiérrez-Dávila et al., in prep).

\begin{figure}
    \centering
    \includegraphics[width=\columnwidth]{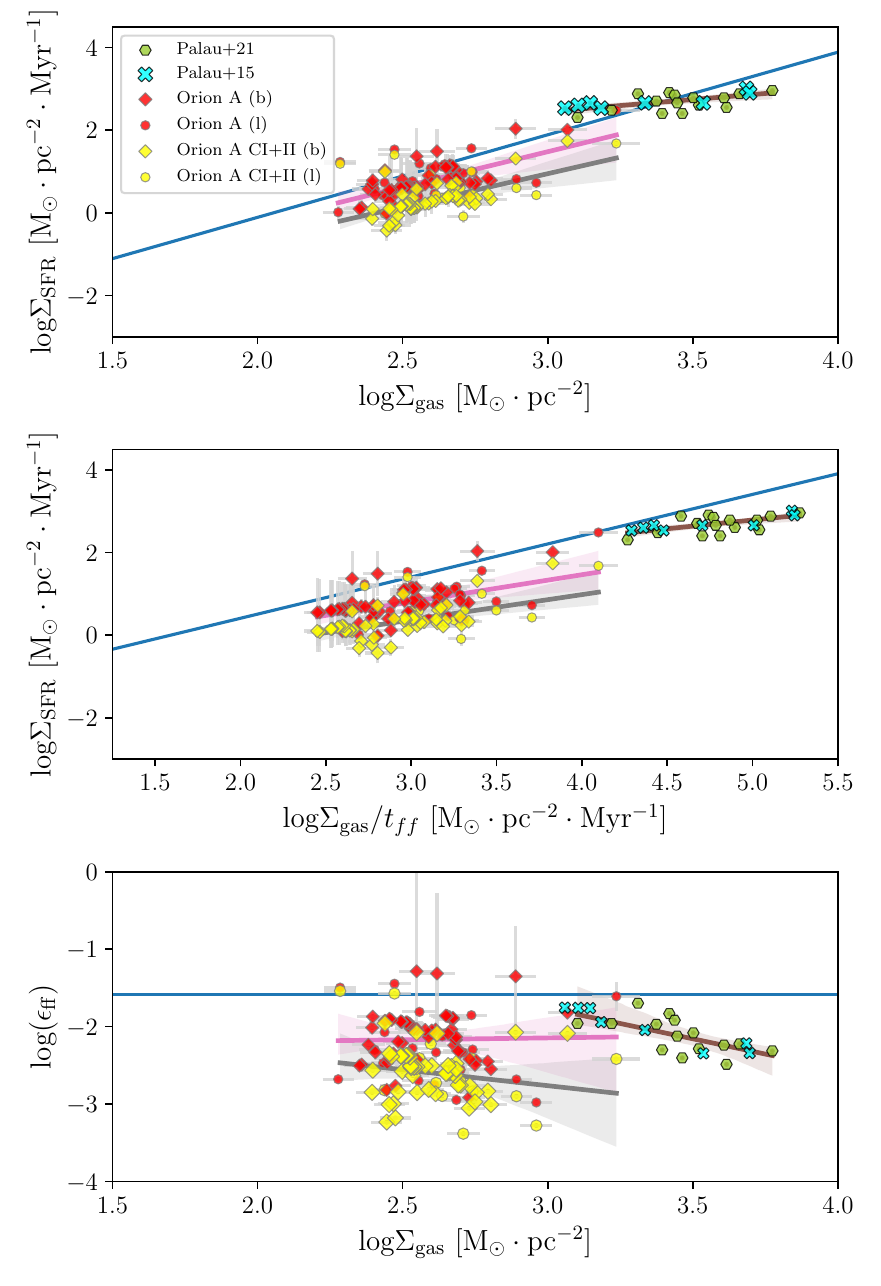}
    \caption{Data point in the three panels use the same symbols and colors for Orion A and Palau et al. data used for Figure \ref{fig:McMyso_3cD}. Top panel: correlation between $\log{\Sigma_{\rm SFR}}$ and $\log{\Sigma_{\rm gas}}$. The blue line corresponds to a fit function of the form $y=2x-4.11$. Center panel: correlation between $\log{\Sigma_{\rm SFR}}$ and $\log{(\Sigma_{\rm gas}/\tau_{\rm ff})}$. The blue line shows a fit of the form $y=x-1.59$. Bottom panel:  $\epsilon_{\rm ff}$ {\it vs.} $\Sigma_{\rm gas}$. The blue line indicates a constant value of $\log \epsilon_{\rm ff}=-1.59$. In all panels, the purple and red lines with dispersion bands show robust linear fits to the data points, detailed in $\S$ \ref{s:results:ss:sfrsgas}}
    \label{fig:SgasPlots}
\end{figure}

In the top panel of Figure \ref{fig:SgasPlots} we show our version of the intra-cloud KS relationship in a $\Sigma_{\rm SFR}$ {\it vs.} $\Sigma_{\rm gas}$ plot that contains data for the Orion A data used in this study (both for Class I+II counts and all YSOs from Class I to III; yellow and red data points, respectively) as well as data from the studies of \citetalias{Palau+15} and \citetalias{Palau+21}. The blue line corresponds to $\log{\Sigma_{\rm SFR}}=2.0\log{\Sigma_{\rm gas}}-4.11$, as in \citetalias{Pokhrel+2021}. The colored lines with dispersion bands correspond to a robust linear fit\footnote{we used the Python routines \texttt{sns} from the \texttt{seaborn} package \citep{sns21} and \texttt{stats} from the \texttt{scipy} package \citet{scipy20}, with a 95\% confidence interval and 1000 bootstrap repetitions.} applied to the Orion A and the Palau et al. points, respectively; for Orion A we obtain slopes of $1.40\pm0.274$ and $1.63\pm0.230$ for the Class I+II and Class I to III YSO count cases, respectively, in disagreement with the expected quadratic fit of \citetalias{Pokhrel+2021}. The situation becomes even more extreme for the data of Palau et al., for which we obtain a slope of $0.56\pm0.161$. Notably, \citet{rawat25} report a slope of 1.46 for their analysis that takes into account clumps in different clouds that have formed bound star clusters.


In the mid panel of Figure \ref{fig:SgasPlots} we plot $\Sigma_{\rm SFR}$ against $\Sigma_{\rm gas}/\tau_{\rm ff}$ for the same dataset as in the top panel. We show a linear fit of the form $\log{\Sigma_{\rm SFR}}=1.0\log{\Sigma_{\rm gas}}-1.59$, as shown in \citetalias{Pokhrel+2021}. Once more, the data for Orion A and those of \citetalias{Palau+15} and \citetalias{Palau+21} have a shallower trend than that of \citetalias{Pokhrel+2021}. When applying a robust linear fit to these datasets, we obtain slopes of $0.76\pm0.124$, $0.75\pm0.131$ and $0.40\pm0.108$, for the Orion Class I+II counts and Class I to III cases and the datasets of Palau et al., respectively, in disagreement with \citetalias{Pokhrel+2021}. In the study of \citet{rawat25} for cluster forming clumps they report a slope of 0.8 for the $\Sigma_{\rm SFR}$ {\it vs.} $\Sigma_{\rm gas}/\tau_{\rm ff}$ relation. 

In the bottom panel of Figure \ref{fig:SgasPlots} we show the behavior of the $\epsilon_{\rm ff}$ as a function of $\Sigma_{\rm gas}$. In the figure, the horizontal blue line at $\log{\epsilon_{\rm ff}}=-1.59$ represents the median value found by \citetalias{Pokhrel+2021} for their sample. Both our Orion A and the \citetalias{Palau+15} and \citetalias{Palau+21} data, fall slightly below the blue line, showing relatively shallow trends. We applied robust linear fits to the Orion A and Palau et al. data, obtaining slopes of $-0.41\pm0.26$ and $-0.02\pm0.25$ for the Orion A Class I+II and Class I to III cases respectively, and $-0.83\pm0.284$ for the Palau et al. data. Despite the shallow slopes, the three datasets show a visible decrease of the  $\epsilon_{\rm ff}$ values at higher surface densities. However, it is also important to notice that, while the data points for the Class I to III case show average values of $\log(\epsilon_{\rm ff})=-2.1$, in the case where we consider only Class I and II sources, the values are smaller, averaging near $\log(\epsilon_{\rm ff})=-2.6$.  For comparison, the Orion A dataset in the study of \citetalias{Pokhrel+2021} agrees roughly with values also close $\log{\epsilon_{\rm ff}}=-2.0$.

\subsection{Mass-Radius relationship \label{s:results:ss:mclreq}}

\begin{figure}
    \centering
    \includegraphics[width=\columnwidth]{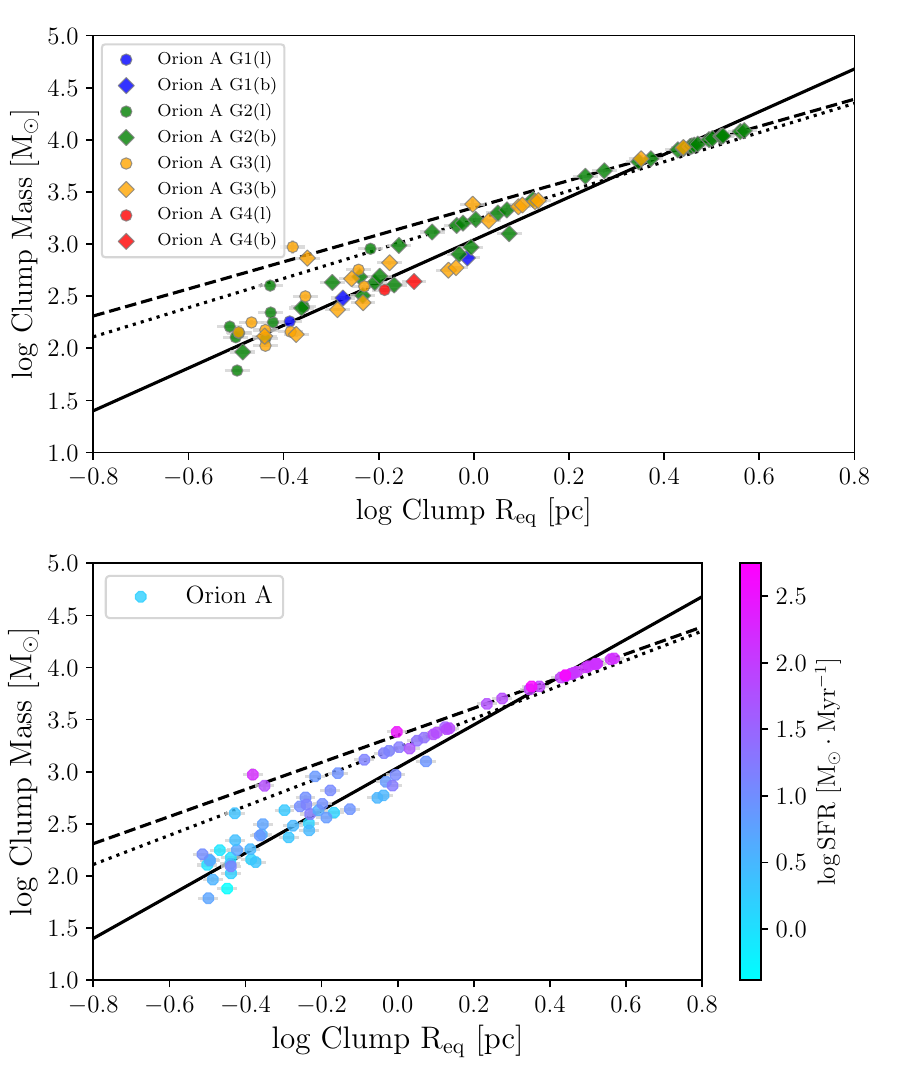}
    \caption{Top panel: $M_{\rm gas}$ {\it vs.} $R_{\rm eq}$ relationship for the Orion A data. The solid, dashed and dotted lines represent data fits with slopes of 2.05, 1.30 and 1.40, respectively. Symbols are colored according to their dendrogram ancestor group and category (branch, leave). Bottom Panel: The symbols are colored according to the clump SFR values. }
    \label{fig:ReqMcl}
\end{figure}

In Figure \ref{fig:ReqMcl} we show the mass {\it vs.} equivalent radius for the objects in Orion A. The dispersion of the data is significantly smaller for this plot than it is for the relations discussed in section \ref{s:results:ss:sfrsgas}. 
Objects with equivalent radii smaller than 1.0~pc exhibit a correlation with a good fit to a function of the form $\log{M_{\rm gas}}=2.05 \log{R_{\rm eq}}+2.44$ (black solid line). The slope close to 2.0 is expected for a collection of structures having nearly the same column density \citep[see Fig. 6 in][]{Lada+08}, which is a good approximation for all structures whose peak $A_V$ value does not rise too high above the threshold value $A_V=7.0$ mag. The plot also shows how a group of structures with the largest mass and radii appear to separate from the group, and are better adjusted with slopes of 1.3 and 1.4. This behavior of a slope flattening for larger radii is found in general in all molecular clouds in the Solar Neighborhood \citep[see, e.g., ][]{Lombardi+10}. 

In the top panel of Fig. \ref{fig:ReqMcl} we show, using colored points, the four common-ancestor groups (G1--G4), each one separated in branches (filled diamonds) and leaves (filled circles). Groups G1 and G4 (blue and red dots respectively), with such few components, have no substantial substructure, but groups G2 and G3 (green and yellow respectively) are split into several branches and leaves. The structures of G2, located at the L1641N-S complex, have the largest projected areas, and thus the lower surface densities, but they also show the largest star-forming rates. This is made clear in the bottom panel, where we color-coded the Orion~A clumps by star forming rate. As it can be seen, most of the clumps with SFR > 1~$M_\odot$/Myr are in the separating branch from the G2 and G3 ancestor groups.

\section{Discussion}\label{s:discsum}

The cycle of star formation (collapse$\leftrightarrow$feedback) is regulated by the stars themselves, affecting the interstellar medium and defining star formation as a highly dynamical phenomenon \citep{Chevance+20}. Basically, scale relations are necessary to guide our understanding of the roles of density segregation, gravitational collapse, and later, feedback that promotes new pilings of gas \citep[arguably the main processes in star formation, e.g.][]{xabiero+12, Ballesteros-Paredes+18, xabiero+19, Alfaro_Roman-Zuniga18, Vazquez+19, RamirezGaleano+22} in different scenarios.

\begin{figure}
    \centering
    \includegraphics[width=1.0\linewidth]{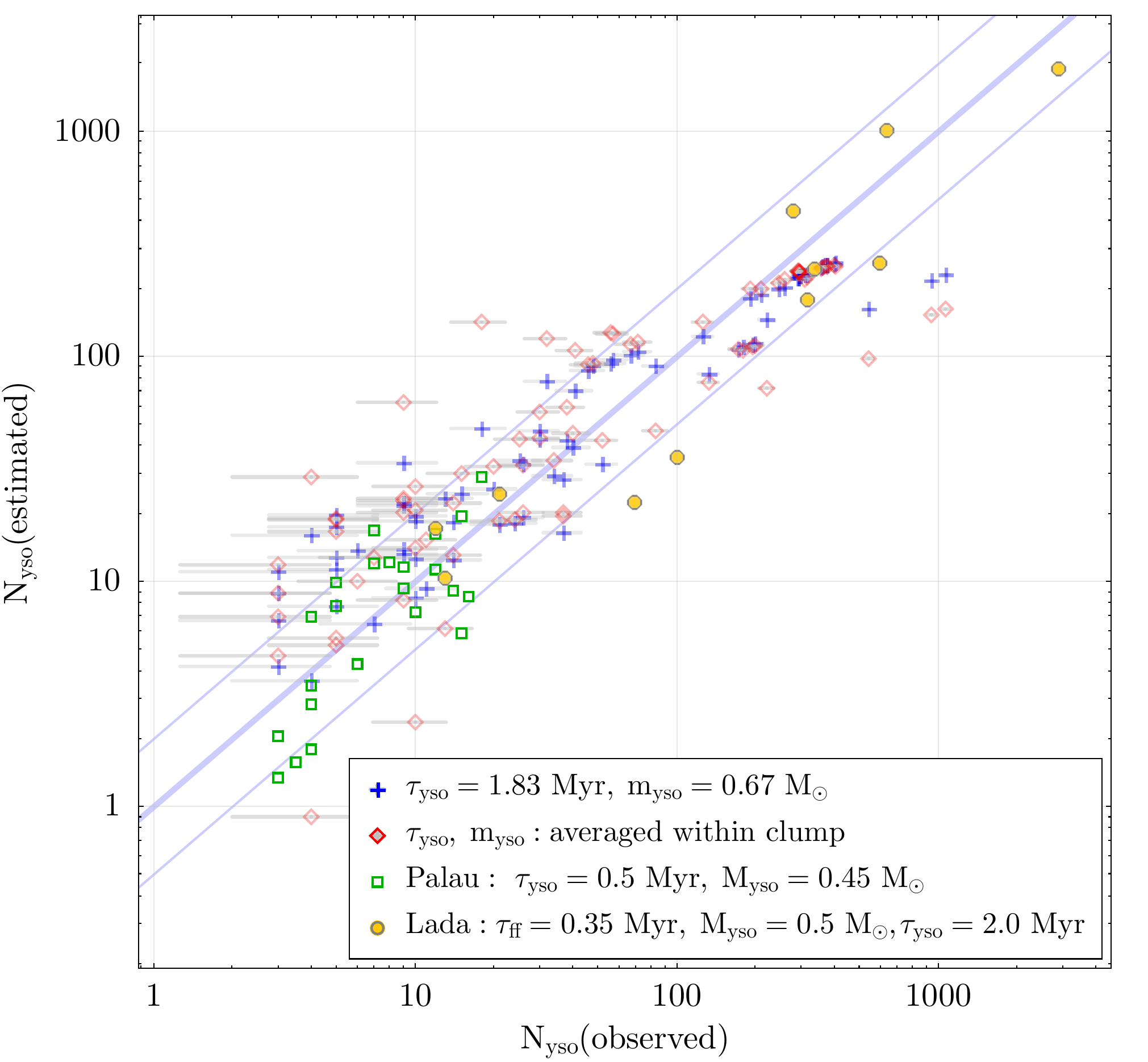}
    \caption{A comparison between observed $N_{\rm yso}$ values and those determined by using eq. \eqref{e15p1}. Blue crosses use constant values for YSO mass and ages and orange diamonds use values averaged within each object in this study. The dark green squares are estimates for the Palau et al. datasets, using constant, typical values for their studies. For comparison, the yellow symbols show estimates for the \citet{Lada+10} data using educated estimates for constant $\tau_{\rm ff}$, $\rm m_{yso}$ and an average yso age of 1.83 Myr. The thick blue line defines the identity, and the two thin blue lines define a factor-of-2 band around it.}
    \label{fig:NysoPred}
\end{figure}

As discussed in \citet{paperI}, the KSR is in fact one way of expressing what they define as a fundamental equation for star formation (FESF), 

\begin{equation}
    \langle\mathrm{SFR}\rangle =\epsilon_{\rm ff}\left (\frac{M_{\rm collapse}}{\tau_{\rm ff}}\right )
\end{equation}
The authors called it fundamental because, as they showed in \citet{paperI}, many different global correlations at galactic or extragalactic scales can be inferred from this relation that expresses the balance between collapse and gas removal by stars.


The regions of GMCs that reach a threshold density, define a mass of gas under collapse, that is converted partially into stars within a time scale that defines that collapse and according to an ``efficiency". This efficiency, in fact, relates the time it takes for a cloud to collapse and the time that takes for the same cloud to exhaust the available gas on the process,  $\epsilon_{\rm ff}= \tau_{\rm ff}/\tau_{\rm depl}$. 

The FESF expresses that those portions of the cloud segregated by density form a number of stars that is proportional to the amount of gas they contain, as it is directly expressed by  $N_{\rm yso}$ {\it vs.} $M_{\rm gas}$, and both are the result of gravitational collapse. What is somewhat surprising, however, is that this relation keeps linearity over a dynamic range that covers the intra-cloud and the inter-cloud cases (i.e., from $\sim 0.1$ pc to $\sim 100$ pc.)

The rough estimates used in Fig. \ref{fig:McMyso_3cD} suggest that the dynamical range could extend to larger scales. The spiral arm scale trend shown in the works of \citetalias{Peltonen+24} and \citet{Zak+25} are of course still subject to discussion, as the scales they probe in the M33 and M51 galaxies are still far from reaching the precision we obtain in Galactic studies, even with state-of-the-art instrumentation. For instance, \ \citetalias{Peltonen+24} infer their $N_{\rm yso}$ values using an indirect method based on inferring total numbers of YSO from the \citet{Kroupa01} IMF of the initial mass function --as only the brightest fraction of the population can be detected--, instead of the direct YSO counts we are able to perform in solar neighborhood star-forming complexes. We can especulate that, if results like those of \citetalias{Peltonen+24} can be confirmed --with reduced scatter-- in the near future,  we could consider the $N_{\rm yso}$ {\it vs.} $M_{\rm gas}$ to be a well defined scaling relation for molecular clouds, with important implications. For instance, this would express the self-similarity of the star formation process we observe at different spatial scales, and it would give strong support to the concept of a FESF and the derivation of other scaling laws from it. 


In order to understand the linear nature of the $\mathrm{N_{\rm yso}\ vs.\ M_{\rm gas}}$, we used equation \eqref{e15p1} to estimate $\rm N_{yso}$ from distinct assumptions in the four parameters of the term in parentheses $(\epsilon_{ff}\langle\tau_{yso}\rangle/\langle m_{yso}\rangle\tau_{ff})$. 

In order to understand the linear nature of the $\rm N_{yso}$ vs $\rm M_{gas}$, we plot in Fig. \ref{fig:NysoPred} the number of protostars counted in the different surveys datasets --$\rm N_{yso}$(observed)-- against the estimated number of protostars that eq. \eqref{e15p1} will predict --$\rm N_{yso}$(estimated)--. In all cases we used a value of $\rm \epsilon_{ff}=0.0078$, corresponding to the median value of $\epsilon_{ff}$ for YSOs of Class I to III in Orion A (see Fig. \ref{fig:SgasPlots}). For each dataset we made different assumptions for the other parameters:

\begin{enumerate}
    
\item We estimated $\rm N_{yso}$(estimated) in Orion A using the $\mathrm{M_{\rm gas}\ and\ \tau_{\rm ff}}$ values from eqs. \eqref{e2.2} and \eqref{e2.5}, but using $m_{\rm yso}=0.67\ M_\odot$ and $\tau_{\rm yso}=1.83$ Myr, which are the mean values of the whole YSO list. These estimates are shown with blue cross symbols in the figure.

\item In a second case, we compute the mean value of $m_{\rm yso}$ and $\tau_{\rm yso}$ in each core, and plug them into eq. \eqref{e15p1} along with $\mathrm{M_{\rm gas}\ and\ \tau_{\rm ff}}$ values from eqs. \eqref{e2.2} and \eqref{e2.5}. These correspond to the red diamond symbols in the figure.

\item We also show the result of the same exercise for the data of Palau. We considered an average mass per object of $\mathrm{m_{yso}=0.45\ M_\odot}$, corresponding to mean mass of stars in the solar neighborhood\citep{IMF24}. We used a typical YSO age of $\tau_{\rm yso}=0.5$ Myr \citep[e.g.][]{Evans+09, Heiderman+10, Lada+10, Dunham15}, and $\tau_{\rm ff}$ obtained from eq.~\eqref{e2.5} using the gas density from \cite{Palau+15} and \cite{Palau+21} for each clump (which averages to $\sim$50000~yr).

\item For comparison, we also made estimates for the data of \citet{Lada+10}, considering, based on their text, average values for $\rm m_{yso}$ and $\tau_{\rm yso}$ of 0.5 $\rm M_\odot$ and 2.0 Myr, respectively. In the case of $\tau_{ff}$, they report an average of 0.35 Myr. We took into consideration the HP2/NICEST mass correction factor of 1.55 discussed in $\S$\ref{s:datamethods:ss:avmaps}.
\end{enumerate}

The figure shows how most of the data points show a good agreement within a factor of 2.~For Orion A, when we move from case (i) to case (ii) we recover the observed $\rm N_{yso}$ with a comparable level of scatter (0.18 RMS) as when using the average $\rm m_{yso}$ and $\rm \tau_{yso}$ from each individual core (0.25 RMS), indicating that those values are not the most crucial for the linearity. 
For the objects in the Palau et al. samples (iii), they are well aligned with the one-to-one relation; this is because the shorter $t_{yso}$ of 0.5 Myr is compensated by the fact that the $t_{ff}$ in these samples is shorter by a similar factor.
For the inter-cloud case of \citet{Lada+10} in (iv), there is a very good
agreement with the identity, but we have to be careful at considering that these data points lie in the middle of large uncertainties, as they consider large objects compared to the intra-cloud sets.

Therefore, the offset discrepancies of Fig. \ref{fig:McNyso_main} appear to vanish using the FESF. This is crucially related to a similarity in the $\rm \tau_{yso}/\tau_{ff}$ ratio (for example, this ratio is very similar for the Lada and Palau samples). A closer agreement with the identity line can be also obtained by using a more precise value of $\rm \epsilon_{ff}$ for each core, which in turn depends on $\tau_{ff}$. The $N_{\rm yso}\ {\it vs.} M_{\rm gas}$ relation is in this sense, a simple but useful dial that summarizes how well we can resolve the star forming capacity of a cloud, independently of the physical scale, depending on how well we understand the relation between the times for collapse and dispersal of gas, and hence the local efficiency of star formation.

Our analysis in $\S$\ref{s:results:ss:sfrsgas} show a disagreement with the mean result of \citetalias{Pokhrel+2021}, with our data showing a shallower slope than quadratic exponent (specially at higher $\Sigma_{\rm gas}$), when considering individual structures at the intra-cloud scale. This could be due to two causes: On the one hand, the portion of the column density under consideration (individual dendrogram objects enclose, in average, higher values compared to using the entire area of the cloud). On the other hand, the methodology of \citetalias{Pokhrel+2021} is essentially different from ours: while we use dendrograms\footnote{The dendrogram analysis considers an object (leave or branch) as a connected set of pixels whose emission is above a given threshold. A different structure is considered if, for a different threshold, the original structure is split into two or more disconnected set of structures.}, \citet{Pokhrel+2021} puts together all detectable regions of dense gas above a given threshold, connected or not, having stars or not, as a single object to define a global behavior. Our analysis, instead, divides the star forming regions into smaller portions, and account only for those clumps that have stars, producing a different star formation relation.  

It is particularly important to consider how the intra-cloud analysis changes the general interpretation for $\epsilon_{\rm ff}$. Our datasets show significant scatter, with $\epsilon_{\rm ff}$ presenting a mean dispersion of 0.5~dex over almost 1.5~dex with a visible decrease at higher $\Sigma_{\rm gas}$ values; this shows that the efficiency of star formation at the intra-cloud level is not constant. Given that the SFR varies for every individual clump in our analysis, it is logical to have a larger scatter.

It is also important to consider the values of star formation efficiency determined directly from the total mass of YSOs and gas, as:
\begin{equation}
    \mathrm{SFE} = \frac{M_{\rm yso}}{M_{\rm yso}+M_{\rm gas}}
\end{equation}

\begin{figure}
    \centering
    \includegraphics[width=\columnwidth]{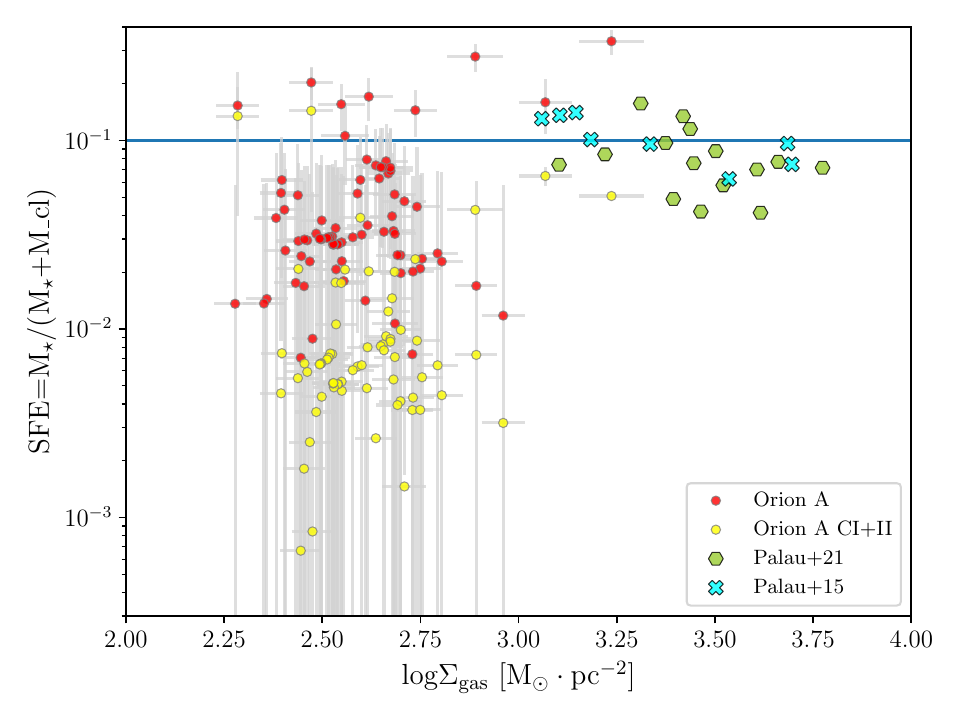}
    \caption{ Star formation efficiency as a function of gas surface density, comparing our Orion A data with those of \citet{Palau+15,Palau+21}. The blue line corresponds to a 10 percent efficiency.}
    \label{fig:sfe}
\end{figure}

In Fig.  \ref{fig:sfe}, we show the SFE for Orion A (using both the cases with and without Class III counts; red and yellow points, respectively) as well as for \citetalias{Palau+15} and \citetalias{Palau+21} (crossmark and hexagon symbols). We observe that the range of SFE values for the case with Class III sources are similar to those of Palau et al., despite being estimated in two very different gas surface density regimes: most points in both datasets lie below the 10 percent efficiency line as expected from previous studies. This is similar to what is obtained from simulations in \citet{paperII}. Notice that the SFE values for the Palau et al. data are, in average larger than those of Orion A, suggesting an increase of SFE at larger $\rm \Sigma_{gas}$ regimes, in line with models by \citet{Gutermuth+11}. The plot suggests that the conversion of gas to stars follows a self-similar process in any region of \textit{dense gas}, independently of the clump size scale and of the surface density of gas. 

The YSO counts that consider only Class I and II sources result in very low values for SFE in most of the points, indicating that it is necessary to consider all sources formed in a region in order to obtain a proper estimate of the global star forming efficiency. Palau et al. achieve this by obtaining their YSO counts at regions with higher density thresholds, which in turn trace earlier stages of formation, and they are closer to detect a majority of the recently formed sources \citep[this is evidenced by the small differences between source counts in mm and IR wavelengths;][]{Palau+13}. The dispersion of the gas and the stellar groups, is rapid but it is not instantaneous, and a region provides information about the gas-to-star conversion process as long as $\epsilon_{\rm ff}$ can be defined.

Following \cite{Grudic+19}, the star-forming efficiency is always small (say, a few tenths of a percent) in clouds at early stages of evolution, while higher values (on the order of 10-30 percent) are measured in later stages, right before gas dispersal is onset. This has been confirmed in cluster forming clumps across diverse clouds \citep{rawat25}. The quantity $\epsilon_{\rm ff}$, can be understood in similar terms. Nonetheless, if expressed as the ratio of the characteristic timescale for star formation collapse to the gas depletion timescale, a large value implies that the available gas mass is always large compared to the collapsed gas mass, independently of the cloud size. Before star formation begins in a cloud, the timescale implied is large, and will diminish as the stellar feedback effects increased; ultimately, the timescale $\tau_{\rm ff}$ vanishes when the parental gas is removed. The period for gas removal in the presence of massive stars, is comparable to the timescale of disk dispersal \citep[stars moving from Class I to Class III evolutionary stages, e.g.][]{Ybarra+13, Roman-Zuniga+15}, but can be larger in the absence of a massive star \citep[e.g.][]{Sandily+15}. 

The collapse timescale is, undoubtedly, a crucial value that defines the rates at which a gas cloud (or clump) can form stars. However, the collapse timescale is in turn governed by the volume of dense gas. The reservoir of gas in a cloud can be maintained by accretion from the exterior \citep[e.g][]{Liu+2015,Olguin+23,Xu+24} and that reservoir will be depleted on a rate that depends directly on the massive stellar content and hence the initial mass function (e.g. \citet{Zamora+12,Zamora+14}; \citet{paperII} and references therein). However, in either case, the star formation will occur at those regions that surpass the threshold density, and the process of star formation will be similar, except that the collapse timescale will be just shorter or longer depending on the volume of dense gas. This is one more reason to why the $N_{\rm yso}$ {\it vs.} $M_{\rm gas}$ relation appears to work in such a large dynamic range.

Another important scaling relation is the mass {\it vs.} size relation that we presented for our intra-cloud analysis in section \ref{s:results:ss:mclreq}. Following \citet{xabiero+12}, clouds or clumps with similar column densities must follow $M\propto R^2$, while branches and leaves within a single cloud must necessarily exhibit a flatter mass-size relation as one moves from low column densities (branches) to leaves (high column densities). Our results show, as expected, that most structures follow nearly a $M-R^2$ relationship,indicating that most of them have a similar average column density. At the same time, the relation flattens out for the largest structures in the L1641 region. Intra-cloud studies like those of \cite[][]{Kauffmann+10a,Kauffmann+10b} showed that the slope in the mass-radius relation moves from 2 to about 1.6-1.7 for the largest structures in nearby clouds, and how the cluster forming portions of clouds ar in better agreement with a slope of the order of 1.3 to 1.5. This is in agreement with independent measurements of the relation on massive gas clumps \citep{Urquhart+18,Xing+22}. \citet{xabiero+12} (see also \S3.3 in \citet{paperI}), showed that the slope of the intra-cloud mass-size relation is directly related to the slope of the column density PDF, since the number of pixels in the range $(\Sigma, \Sigma+\delta\Sigma)$ that define the PDF define the area, and thus, the equivalent radius of the cloud. Thus, one can argue that, since the column density PDF is directly related to the mass-size relation, and thus, to the mass and the area of the cloud, the KS relation is somehow defined by the column density PDF of molecular clouds.

\section{Summary}\label{s:sum}

In this paper, we present a simple analysis around scaling relations derived from the Schmidt conjecture for a giant star-forming molecular cloud at the intra-cloud scale. For this purpose, we used a column density map and a catalog of young stars.

Using a hierarchical tree (dendrograms), we separated coherent gas structures in the column density maps, above a constant threshold defined by $A_V=7.0$ mag, which is known to define the areas of clouds containing dense molecular gas that can actively form stars. We carefully counted the number of YSOs projected within the boundaries of each object, to have an estimate of their current embedded population. We then calculated both the gas and star formation surface densities for each structure, as well as their free-fall timescales. 

Using these data, we constructed plots that show the behaviour of our data at the intra-cloud level in three known scaling relations, (a) $N_{\rm yso}$ {\it vs.} $M_{\rm gas}$ (along with a $M_{\rm yso}$ {\it vs.} $M_{\rm gas}$); (b) $\Sigma_{\rm SFR}$ {\it vs.} $\Sigma_{\rm gas}$ and (c) $R_{\rm eq}$ {\it vs.} $M_{\rm gas}$. We also showed the behaviour of $\epsilon_{\rm ff}$ {\it vs.} $\Sigma_{\rm gas}$, which according to the study of \citetalias{Pokhrel+2021} is expected to be approximately constant across single clouds. 

Our data show that the gas structures traced by dendrograms in the three clouds show a linear behavior, in line with inter-cloud versions of the $N_{\rm yso}$ {\it vs.} $M_{\rm gas}$ by \citetalias{Lada+10} and \citet{Jorgensen+08}, as well as with the individual dense clump determinations by \citetalias{Palau+15} and \citetalias{Palau+21}. This suggests that the relation is valid for a dynamic range in gas mass of more than 3 orders of magnitude from $10$ to $10^4$ M$_\odot$. We also especulate that, by considering estimates from the study of \citetalias{Peltonen+24} for clouds in the M33 galaxy, the relation could possibly be valid over 5 orders of magnitude. 

We used our data to construct plots of $\log \Sigma_{\rm SFR}$ against both $\log \Sigma_{\rm gas}$ and $\log{(\Sigma_{\rm gas}/\tau_{\rm ff})}$. The datasets we used are better fit by shallower slopes than those of \citetalias{Pokhrel+2021}. We also constructed a $\log \epsilon_{\rm ff}$ {\it vs.} $\log \Sigma_{\rm gas}$, which showed relatively good agreement with a constant average value across the clouds, but with larger scatter and an apparent decrement at large surface densities of gas.

Our $R_{\rm eq}$ {\it vs.} $M_{\rm gas}$ shows a quadratic relation for most of the low-mass, low-SFR structures, with a shallower slope of the order 1.3-1.5 for the largest structures with higher values of local SFR. Such behavior has also been observed in previous works, and possibly reflects how the density distribution is different in large structures that form clusters of stars and not only small groups.

In all of our results, it is clear that the free-fall timescale of the star-forming gas is a crucial parameter, which in turn defines the volume of the dense gas region collapsing to form stars. This timescale is probably larger at earlier stages, when gas is being accreted towards the gas clumps, but will shrink after the onset of stellar feedback, when the available reservoir of gas is depleted.  As discussed in \citet{paperI} and \citet{paperII}, the efficiency of the gas-to-star conversion depends crucially on the balance between the amount of gas reaching the threshold density for collapse, and the removal of gas by the action of the star that just formed. The value of the star formation efficiency per free-fall time, according to our results, is of the order of a few percents, and it has a small dependency on the gas surface density. 

\section*{Acknowledgements}

C. R.-Z., J.B.-P. and A.P. acknowledge support from project UNAM-DGAPA-PAPIIT IG101723 and from CONAHCyT project number 86372 of the ‘Ciencia de Frontera 2019’ program, entitled ‘Citlalcóatl: A multiscale study at the new frontier of the formation and early evolution of stars and planetary systems’, Mexico. A.P. acknowledges financial support from the UNAM-PAPIIT IN120226 grant, and the Sistema Nacional de Investigadores of SECIHTI, M\'exico.

This research made use of \texttt{astrodendro}, a Python package to compute dendrograms of Astronomical data (http://www.dendrograms.org/). This work made use of \texttt{Astropy}:\footnote{http://www.astropy.org} a community-developed core Python package and an ecosystem of tools and resources for astronomy \citep{astropy+2013, astropy+2018,astropy+2022}.

Dynamic visualization of the different datasets used in this paper was crucial for our analysis. Such task, along with the construction of two of our figures, were performed using TOPCAT \citep{topcat05}, an essential tool for Astronomy.

In the present paper we discuss observations performed with the ESA Herschel Space Observatory \citep{HSA10}, in particular employing Herschel's science payload to do photometry using the PACS \citep{PACS10} and SPIRE \citep{SPIRE10} instruments.

 We acknowledge Claire Dobbs for providing estimates of $N_{\rm yso}$ {\it vs.} $M_{\rm gas}$ from her 2022 simulations under personal request.

\section*{Data Availability}

The YSO catalog for Orion A is available from a public repository, as described in the cited source. The HP2 map is already available in a public repository, as described in the corresponding cited sources. We can provide access to the dendrogram object catalogs we obtained from the HP2 map, upon reasonable request. The additional data used in figures 3-6 and 8, were obtained by personal request to the authors of the corresponding cited sources (Joshua Peltonen, Aina Palau and Claire Dobbs). 



\bibliographystyle{mnras}
\bibliography{references01} 

\appendix

\appendix

\section{Extragalactic Contamination} \label{App.1}

\begin{figure*}
    \centering
    \includegraphics[width=1.9\columnwidth]{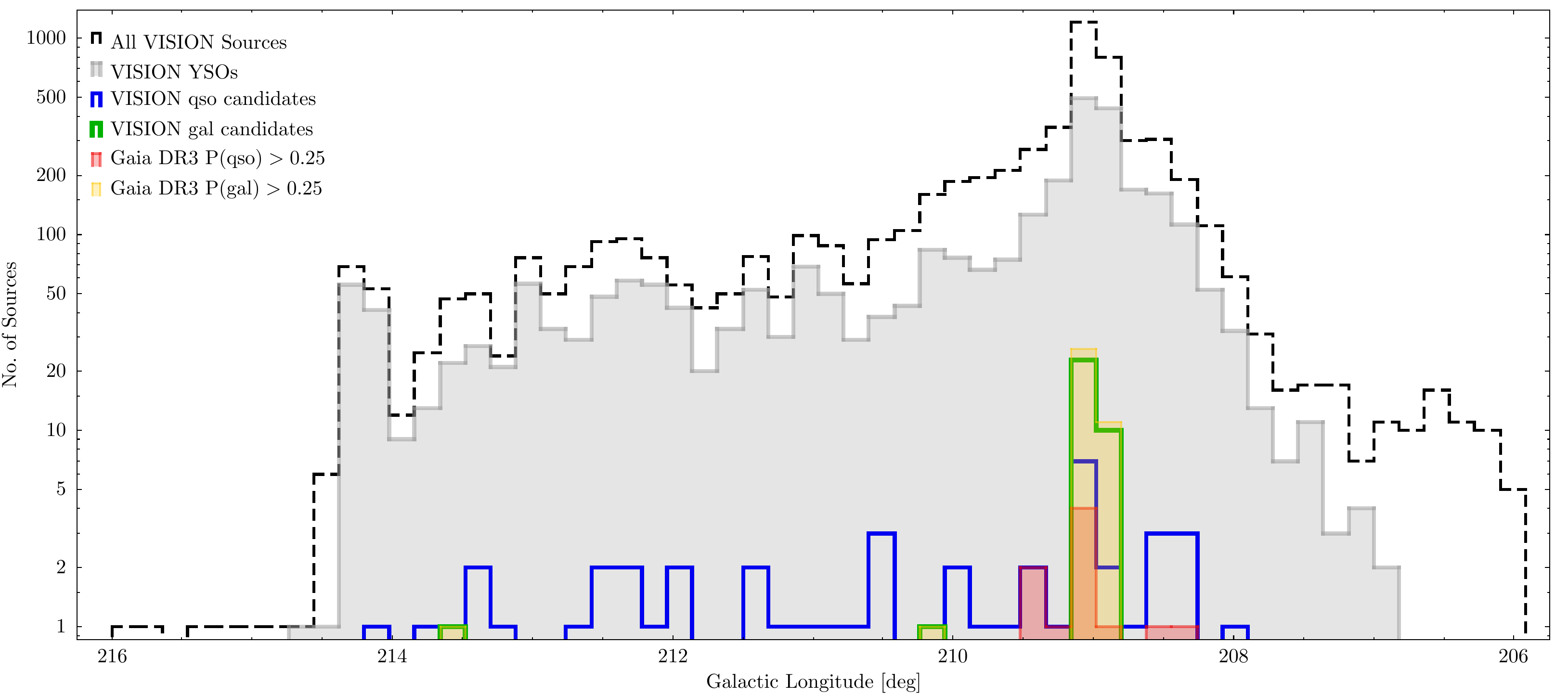}
    \caption{Distribution of sources in the VISION catalog as a function of Galactic Longitude. Sources catalogued as candidates for quasar and galaxy contaminants with VISION labels (\citetalias{Grossschedl+19}) are represented by the green and blue histograms. Sources catalogued with the DSCCM criteria of Gaia DR3 \citep{Jamal24} are represented with the red and yellow histograms.}
    \label{fig:qsogal}
\end{figure*}

In Fig. \ref{fig:qsogal} we show the distribution of Orion A VISION sources as a function of Galactic Longitude. The notorious peak near $\mathrm{209^\circ}$ corresponds roughly to the Integral Field and the ONC. Extragalactic flags (quasars, field galaxies) are present in about 3 percent of the YSO list; quasars appear to be relatively well distributed across the whole cloud, while confusion with field galaxies is mostly relevant for the ONC region. We also show the distribution of sources with Gaia DR3 DSCCM $\mathrm{P_{qso,gal}>0.25}$, which appear to be more biased towards the ONC region. We opted for a conservative decision and removed all VISION candidates from our YSO counts. We confirmed than this selection, in turn, removes all sources with $\mathrm{P_{qso}>0.025}$ and $\mathrm{P_{gal}>0.25}$.

\section{Detail on YSO ages} \label{App.3}

Here we provide a more detailed justification for the choice of 2 and 3 Myr ages for Class II (cTTS) and Class III (wTTS) YSOs, respectively, in our analysis.

For Orion and particularly for the ONC, \citet{DaRio10} found a 0.3-0.4 dex age spread for T Tauri stars, with a mean age of 3-4 Myr. \citet{Beccari17} found that the ONC probably has three distinct populations defining three pre-main sequence groups, the youngest being the most spatially concentrated. The mean ages of those three groups are 1.2, 1.9 and 2.9 Myr. Now, Focusing on studies that tried to make specific separate statistics for classic and weak T Tauri stars (although not necessarily in Orion): 

\citet{Galli15} studied a sample of T Tauri stars in Lupus. They found model dependent differences in ages (they compared ages derived using \citet{DAntona97}, \citet{Baraffe98} and \citet{Siess00} models) but the same tendency for a clear separation in the mean and median ages for cTTS and wTTS: Mean and median ages for cTTS range from 1.4 to 2.4 Myr and 1.2 to 2.2 Myr, respectively, while mean and median ages for wTTS range from 6.2 to 11 Myr and 2.6 to 6.3 Myr, respectively.

\citet{Grankin16} pointed out that PMS stars with masses 0.3-1.1 M$_\odot$ occupy the same region in the HR diagram, within the 1 and 10 Myr isochrones, and their mass and age distributions are very similar, both bimodal. They report a wTTS mean age of approx 2.5 Myr and a cTTS mean age close to 2 Myr, with a suggestion for a disk dissipation timescale of 0.5 Myr in average. Importantly, they claim that the wTTS are approx 1 Myr older than the cTTS in their sample.

\citet{Hernandez23} made a careful analysis of cTTS, wTTS and an intermediate “cwTTS” (end of accretion phase) sample across a large portion of the Orion OB1 region, and compared ages derived from spectral properties and isochrones from the PARSEC \citep{PARSEC} and MIST \citep{MIST0} evolution models. With PARSEC they found median ages of 2.8 , 3.7 and 5.9 Myr for cTTS, cwTTS and wTTS, respectively. With MIST models they found median ages of 2.4, 3.1 and 3.7 Myr for cTTS, cwTTS and wTTS, respectively.

Based on the studies mentioned above, which use a good diversity of samples, methods and evolutionary models, we considered adequate to use ages of 2.0 and 3.0 Myr for Class II and Class III sources our revised analysis.


\bsp	
\label{lastpage}
\end{document}